\documentclass[12pt,a4paper,english]{article}
\usepackage[T1]{fontenc}
\usepackage[latin1]{inputenc}
\usepackage{float}
\usepackage{graphicx}

\makeatletter

\usepackage{graphicx}

\usepackage{babel}
\makeatother
\begin{document}
\begin{center}November, 2005\\
 \begin{large}
{\textbf{Hybridization functions of intermetallic Ce compounds with group VI elements}}\\
\vspace{.4cm}
P. Roura-Bas~\(^{a,*}\) ,V. Vildosola~ \(^{a,b}\)~~and A. M. Llois~\(^{a,b}\)\\
\end{large} \vspace{0.2cm} \begin{small}
{\textit{\(^a\) Departamento de F\'{i}sica, Centro At\'{o}mico Constituyentes, Comisi\'{o}n 
Nacional de Energ\'{i}a At\'{o}mica}}\\
{\textit{\(^b\) Departamento de F\'{i}sica, Universidad de Buenos Aires, Buenos Aires, Argentina}}\\
\end{small}\end{center}

\vspace{1cm}
\begin{small}
Among the Cerium compounds, the heavier monochalcogenides CeY (Y=S,Se,Te) of cerium 
provide a class of compounds which together with the heavier monopnictides CeX (X=N,P,As,Sb,Bi) 
exhibit extremely interesting physical properties. The hybridization of the partially delocalized 
f-electrons seems to be responsible  for the unusual properties, in particular the magnetic ones 
which show great chemical sensitivity.
As compared to the monopnictides, the monochalcogenides have an additional anion p-electron 
which shifts the Fermi energy into the p-region of the Cerium 5d-derived density of states, and which
should 
affect the hybridization function of\index{\footnote{}} the electrons in the conduction band. In this contribution 
we analyze the 4f-hybridization function and the crystal-field splittings of the compounds under 
study. 

\vspace{0.3cm}
$^{*}$ Corresponding author

CAC-CNEA, Av. Gral. Paz 1499, San Mart\'in (1650), Buenos Aires, Argentina

roura@tandar.cnea.gov.ar
\end{small}

\vspace{1cm}

CeY and CeX compounds with Y= S, Se, Te and X= P, As, Sb, Bi are suitable systems to extend the knowledge on the relationship between 
electronic structure and the physical properties of highly correlated Ce compounds, because they 
have a simple NaCl crystal structure. In this systems Ce ions are in a trivalent state with one 4f electron. Due to the cubic crystal fields, the $J=5/2$ sixfold degenerate 
magnetic ground state splits into the $\Gamma_{8}$ excited quartet state and the $\Gamma_{7}$ 
ground state doublet$[1]$. CeS, CeSe and CeTe exhibit antiferromagnetic  transitions
at $T_{N}=8.4$ K, $5.4$ K, and $2.2$ K, respectively while the CeX compounds have antiferromagnetic transitions 
at $T_{N}= 7.1 $K, $7.2$ K and $18$ K respectively $[2]$.
There is a small reduction of the crystal-field splittings on going from CeX series to CeY series.
The splitting energies between the two $J_{5/2}$ states are  considerably reduced for the heavier compounds CeTe, CeSb, CeBi $[3]$. These compounds were studied by N. Kioussis {\it et al} $[4]$ by doing first-principle calculations 
based on the self-consistent, warped-muffin-tin LMTO method, in which only the spherical part of the 
potential is considered
and the $4f$ state is treated as a core-like one. These authors can explain the experimetal crystal-field splittings calculating the hybridization induced splittings on top of the extrapolated point-charge levels $[5]$. 

In this contribution, we study by an alternative method, the decrease of the crystal-field splittings on going from CeX  to CeY series and the influence of having an extra $p$-electron in CeY series as compared to CeX one.
We calculate these splittings using an {\it ab initio}-many body combined technique. 
Within the {\it ab initio} frame the 4f state belongs to the valence band and the hybridization function 
is obtained from the 4f-projected density of states (DOS) following ref. $[6]$. We use the all-electron full-potential 
linearized augmented plane wave (FP-LAPW) method as implemented in the Wien-2k code $[7]$ and the local density approximation (LDA) for the exchange and correlation potential.
The calculations are perfomed for the systems at their experimental values. The {\it muffin-tin} radii, $R_{mt}$ are taken equal to 2.4 a.u. for Ce in all the compounds, and the corresponding radii for the anion ligand varies from 1.6 a.u. to 2.8 a.u. depending on the atomic radius. 102 {\bfseries k} points are considered in the irreducible Brillouin zone.

Due to the extra p-electron of the Y-anion in the monochalcogenides,
the hybridization function shows that below the Fermi energy $(E_{F})$, 
the $\Gamma_{8}$ symmetry  is more strongly hybridized in CeY than in CeX, as it can be drawn from the 
comparison of Fig. 1 and Fig. 2.
Two thirds of the additional electron in CeY go into the insterstitial region and contributes 
to the enlarged $\Gamma_{8}$. The {\it ab initio} obtained hybridization functions are 
taken as inputs to calculate the crystal-field 
splittings within NCA (non crossing approximation) for the infinite-$U$ Anderson model $[8]$.
While in the LDA calculation the $4f$ state is a valence state, in 
the NCA-equations, the {\it bare} value of the  $4f$ state is taken from experimental photoemission experiments. These values are -3 eV for CeX, -2.6 eV for CeTe and -2.4 eV for CeS and CeSe with 
respect to the Fermi level $[3]$. Within NCA  the self energies and the Green's functions 
are self-consistently obtained by considering the spectrum 
up to 6500 K above the Fermi level
for the more delocalized systems (Y=S,Se, X=P,As) and up to 5000 K for the more localized ones 
(Y=Te, X=Sb). The criterion for choosing the considered energy range was discused in another contribution 
to this conference $[9]$. The energy spectrum around the Fermi energy plays an important role in the 
determination of physical properties such as the
crystal-field splittings. We obtain the $\Gamma_{7}$ state as the ground state for 
both series of Ce compounds. The values of the   
calculated crystal field  splittings, $\Delta_{CF}=E(\Gamma_{8})-E(\Gamma_{7})$ , are in very good agreement with the experimental ones, as shown in table 1.
The crystal-field splittings are of the same order of magnitude in 
CeY and in CeX, in fact $\Delta_{CF}$(CeY)'s are slightly smaller than $\Delta_{CF}$(CeX)'s both in our results and the experimental data.
However, looking at the  hybridization functions of the $\Gamma_{7}$ and $\Gamma_{8}$ states and comparing them
with the partial $p$ and $5d$ densities of states shown in the insets of Fig. 1 and Fig. 2,  one would expect 
at first sight a larger splitting for CeY than for 
CeX due to the  much larger $\Gamma_{8}$ hybridization values below $E_{F}$ for the first 
series than for the second one. From the comparison of the partial densities of states it is clear that 
below $E_{F}$, $\Gamma_{8}$ hybridizes essentialy with the  
\begin{small}{\itshape p} \end{small}-states of the X and Y anions, while $\Gamma_{7}$ with the 
\begin{small}{\itshape 5d} \end{small}-states. Our results indicate that  
for these Ce compounds, the $\Delta_{CF}$'s values are determined mainly by the hybridization character near the Fermi level. The extra electron of the CeY 
compounds does not change the symmetry of the ground and excited states preserving the order of magnitude of the 
$\Delta_{CF}$'s when going from CeX and CeY. The slight differences in the 
crystal-field splittings of CeX with respect to those of CeY  could be understood by the fact that the Fermi level falls in a 
maximum of the \begin{small}{\itshape 5d} \end{small}-DOS in all the CeX compounds and in a valley of the 
\begin{small}{\itshape 5d} \end{small}-DOS  in all the CeY cases. Consequently, the $\Delta_{CF}$'s are reduced in CeY as compared to CeX.\\ 
\\ 
This work was partially funded by UBACyT-X115, PICT-0310698 and PIP 2005-2006 Num. 6016. A. M. Llois and V. L. Vildosola belong to CONICET (Argentina).

\begin{table}[a]
\begin{center}
\begin{tabular}{|c|c||c|c|}
\hline
Y & ~~~~~~~~~$\Delta_{CF} ^ {exp}$~~~~~~~~~~~$\Delta_{CF} ^ {NCA}$~~ 
& X &~~~~~~~~~$\Delta_{CF} ^ {exp}$~~~~~~~~~~~$\Delta_{CF} ^ {NCA}$~~\\
\hline
S &~~~~~~~~~~~140~~~~~~~~~~~~150~~ & P &~~~~~~~~~~ 172~~~~~~~~~~~~160~~  \\
\hline 
Se& ~~~~~~~~~~116~~~~~~~~~~~~120~~ & As &~~~~~~~~~~ 159~~~~~~~~~~~~150~~\\
\hline
Te &~~~~~~~~~~~32~~~~~~~~~~~~~50~~ & Sb &~~~~~~~~~~~37~~~~~~~~~~~~~70~~\\
\hline
\end{tabular}
\end{center}
\caption{Crystal-field splittings $\Delta_{CF}$ of CeX and CeY systems. 
Experimental and calculated results are given in Kelvin. See Reference [3] for experimental data.}
\end{table}

\begin{center}
References\\
\end{center}
$[1]$ D. Adroja, {\it et al}, Phys. Rev. B {\bfseries 62} 12181 (2000) \\
$[2]$ Handbook on The Physics and Chemistry of Rare Earths, Vol 15, Chapter 98, 
Elsevier Science Publishers B.V., 1991 \\
$[3]$ M. Nakayama {\it et al}, Phys. Rev. B {\bfseries 69} 155116 (2004); H. Heer, {\it et al}, J. Phys. C: Solid State Phys. {\bfseries 12} 5207 (1979)\\
$[4]$ N. Kioussis {\it et al}, Phys. Rev. B {\bfseries 44} 10003 (1991); B. R. Cooper {\it et al}, Phys. Rev. B {\bfseries 36} 3809 (1987) \\
$[5]$ R. Birgeneau {\it et al}, Phys. Rev. B {\bfseries 8} 5345 (1973)\\
$[6]$ O. Gunnarsson \textit{et al.}, Phys. Rev B \textbf{39} (1989) 1708; J. E. Han \textit{et al.}, Phys. Rev. Lett. \textbf{78} (1997) 939\\
$[7]$ P.Blaha,K.Schwartz,G.K.H.Madsen,D.Kvasnicka and J.Luitz,WIEN2k, An Augmented 
Plane wave + Local Orbitals Program for Calculating Crystal Properties (Karlheinz Schwarz, Techn. 
Universitait Wien, Austria) 2001.ISBN 3-9501031-1-2\\
$[8]$ N. Bickers {\it et al}, Phys. Rev. B {\bfseries 36} 2036 (1987); N. Bickers, Reviews of Modern Phys. {\bfseries 59} 845 (1987) \\
$[9]$ P. Roura-Bas, V. L. Vildosola and A. M. Llois, accepted for publication in Physica B.\\


\begin{figure}[b]
\begin{center}
\scalebox{0.5}{\includegraphics{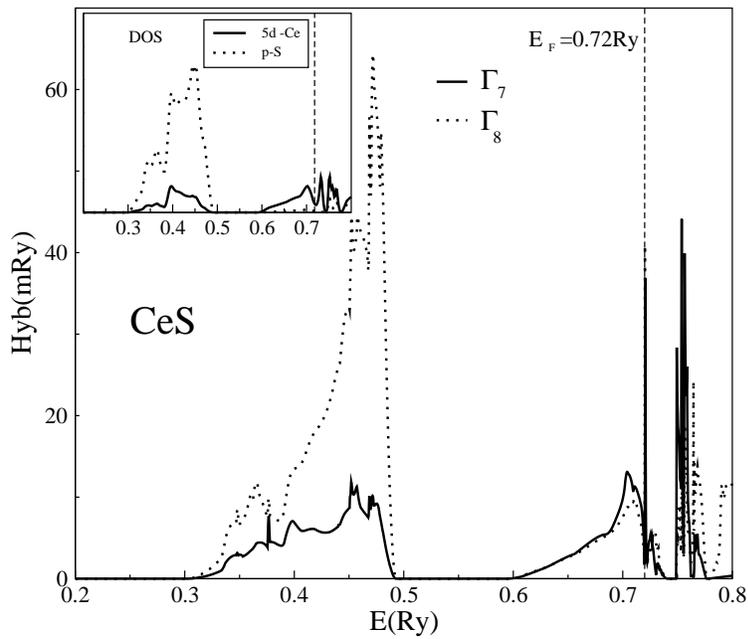}}
\end{center}
\caption{Calculated hybridization functions for CeS: $\Gamma_{7}$ symmetry (solid curve) and $\Gamma_{8}$ symmetry (dotted curve). Inset: $5d-$DOS for Ce (solid curve) and $p-$DOS for the anion S (dotted curve).}
\end{figure}

\newpage
\begin{figure}[c]
\begin{center}
\scalebox{0.5}{\includegraphics{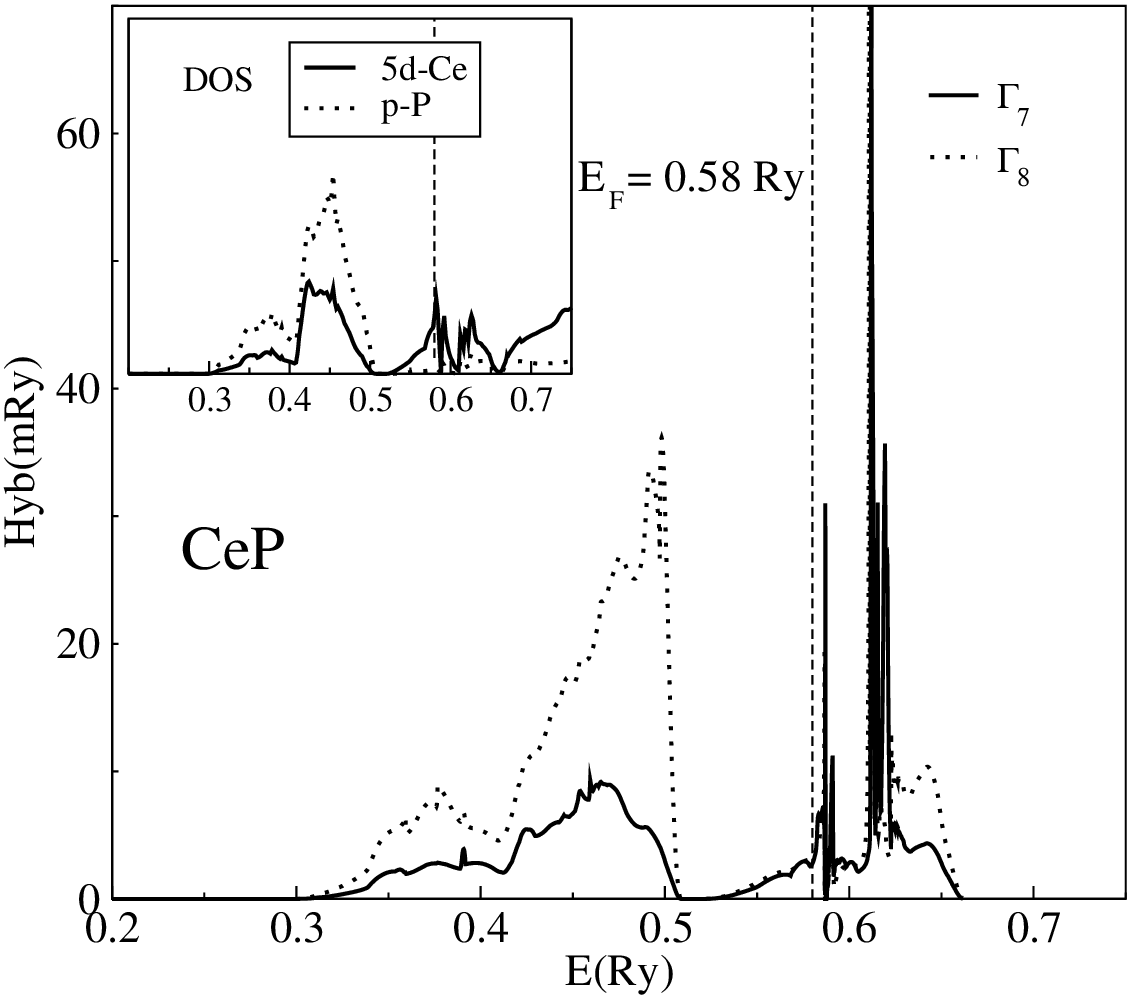}}
\end{center}
\caption{Calculated hybridization functions for CeP: $\Gamma_{7}$ symmetry (solid curve) and $\Gamma_{8}$ symmetry (dotted curve). Inset: $5d-$DOS for Ce (solid curve) and $p-$DOS for the anion P (dotted curve). }
\end{figure}

\end{document}